\documentclass[
  aps,
  prx,
  reprint,
  superscriptaddress,
  longbibliography,
  nofootinbib,
  floatfix
]{revtex4-2}

\usepackage{amsmath,amssymb,bm}
\usepackage{amsthm}
\usepackage{graphicx}
\graphicspath{{../}{./}}
\usepackage{booktabs}
\usepackage{placeins}
\usepackage[unicode,colorlinks=true,allcolors=blue]{hyperref}
\usepackage{orcidlink}
\hypersetup{pdftitle={Resilience of Emergent Hamiltonian quantum memories in the presence of noise},
  pdfauthor={Anubhab Sur and Rubem Mondaini}}

\newcommand{\ii}{\mathrm{i}}
\newcommand{\dd}{\mathrm{d}}

\newcommand{\ket}[1]{\lvert #1\rangle}
\newcommand{\bra}[1]{\langle #1\rvert}
\newcommand{\calN}{\hat{\mathcal{N}}}

\newcommand{\figtwo}[2]{\includegraphics[width=\textwidth]{#1}}

\begin{document}

\title{Resilience of emergent Hamiltonian quantum memories in the presence of noise}

\author{Anubhab Sur\,\orcidlink{0009-0007-0359-865X}}
\email{asur3@uh.edu}
\affiliation{Department of Physics, University of Houston, Houston, Texas 77004, USA}
\affiliation{Texas Center for Superconductivity, University of Houston, Houston, Texas 77004, USA}

\author{Rubem Mondaini\,\orcidlink{0000-0001-8005-2297}}
\email{rmondaini@uh.edu}
\affiliation{Department of Physics, University of Houston, Houston, Texas 77004, USA}
\affiliation{Texas Center for Superconductivity, University of Houston, Houston, Texas 77004, USA}

\begin{abstract}
Emergent Hamiltonians can freeze an entangled state prepared through coherent evolution, since that state is an eigenstate of the Hamiltonian under which it is then stored. We show that the freedom left in this construction allows the storage spectrum to be adapted to a noisy environment without changing the stored state. For a product of Bell pairs prepared from a density wave in an engineered chain, a selection rule prevents local density noise from connecting the target to any of its degenerate partners, so that rescaling the storage Hamiltonian by a factor $\alpha$ moves every transition this noise drives. Under $1/f$ noise, the storage time then grows in proportion to $\alpha$, a result we confirm by evolving the memory under explicit classical $1/f$ fields, whereas for an Ohmic bath the most favorable scale depends on its cutoff. Fermionic loss, treated as a solvable model, includes a transition that rescaling leaves at zero frequency, and an energy shift $\mu$ per particle moves it away from zero, with an exact orbital solution giving the resulting lifetimes. Finally, among number-conserving quadratic reference Hamiltonians, $\alpha$ and $\mu$ are the only controls that keep the storage Hamiltonian nearest neighbor at every time along the preparation, so that the protocol combines local freezing with control over the storage spectrum.
\end{abstract}

\maketitle

\section{Introduction}
\label{sec:introduction}

Preserving an entangled many-body state, a fundamental aspect of quantum technologies, requires keeping both its local properties and its correlations between distant parts of the system. Several approaches address aspects of this problem. For instance, many-body localization is argued to preserve a memory of the initial local occupations in the presence of sufficiently large disorder~\cite{NandkishoreHuse2015,AbaninEtAl2019,SchreiberEtAl2015,SmithEtAl2016}, although this memory is mostly local and is eroded by coupling to an environment~\cite{LeviEtAl2016,FischerEtAl2016,LueschenEtAl2017}. Quantum error correction, in turn, protects encoded information by identifying and correcting errors~\cite{Shor1995,KnillLaflamme1997,Terhal2015,GoogleQuantumAI2025}, while other strategies build protection into the Hamiltonian itself, either by encoding information in decoherence-free subspaces~\cite{ZanardiRasetti1997,LidarChuangWhaley1998} or by separating the protected states from the rest of the spectrum with an energy gap~\cite{BaconBrownWhaley2001,JordanFarhiShor2006,GyenisEtAl2021}. In Ref.~\cite{SurGuoMondaini2026}, we explored a complementary route based on emergent Hamiltonians: by evolving an initially simple state into an entangled state, one can freeze it by switching to a Hamiltonian for which it is an eigenstate. This protocol preserves the complete state under isolated evolution. It allowed us to prepare and store products of maximally entangled Bell pairs and globally entangled Greenberger-Horne-Zeilinger (GHZ) states.

\begin{figure*}[t]
  \centering
  \figtwo{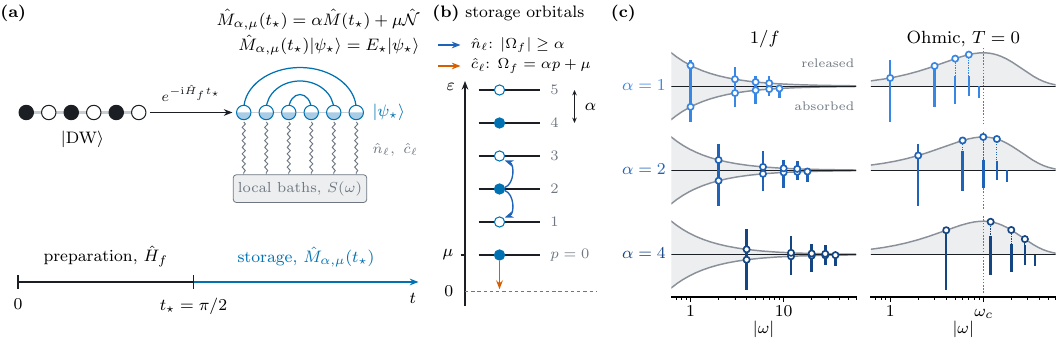}{figures/figure1_v2.pdf}
\caption{Preparing, freezing, and storing a Bell-pair memory. (a) A six-site density wave $\ket{\mathrm{DW}}$ evolves under $\hat H_f$ into mirror Bell pairs $\ket{\psi_\star}$ at $t_\star=\pi/2$. A quench to the storage Hamiltonian $\hat M_{\alpha,\mu}(t_\star)$ freezes this state, which is then coupled to independent local baths through $\hat n_\ell$ or $\hat c_\ell$. (b) One-particle storage orbitals of $\hat M_{\alpha,\mu}(t_\star)$, with energies $\varepsilon_p=\alpha p+\mu$. The Bell target fills the even orbitals. A density operator moves one fermion from an even to an odd orbital, so that $|\Omega_f|\geq\alpha$ (blue arrows). Loss removes a fermion from orbital $p$ and releases $\alpha p+\mu$ (orange arrow), which vanishes for $p=0$ when $\mu=0$. (c) Density-noise transition weights for $L=10$ and $\alpha=1,2,4$. Each bar sums the squared matrix elements over sites and over all transitions out of the target at one signed frequency $\Omega_f$, and is placed at $|\omega|=|\Omega_f|$. Upward (downward) bars denote energy released (absorbed) by the memory. The $1/f$ and zero-temperature Ohmic curves, on separate vertical scales, show the spectra sampled by these transitions. Open circles mark the spectral values at the transition frequencies, and the dotted line marks the Ohmic cutoff $\omega_c=10$.}
  \label{fig:concept}
\end{figure*}

The construction starts with a state $\ket{\psi_0}$ that is an eigenstate of a reference Hamiltonian $\hat H_0$. Evolution under a preparation Hamiltonian $\hat H_f$ produces $\ket{\psi(t)}=\hat U(t)\ket{\psi_0}$, where $\hat U(t)=e^{-\ii\hat H_f t}$. The emergent Hamiltonian~\cite{VidmarIyerRigol2017,VidmarXuRigol2017,ModakVidmarRigol2017, ZhangVidmarRigol2021},
\begin{equation}
  \hat M(t)=\hat U(t)\hat H_0\hat U^\dagger(t)\ ,
  \label{eq:emergent-hamiltonian}
\end{equation}
has $\ket{\psi(t)}$ as an eigenstate with the same eigenvalue as $\ket{\psi_0}$ under $\hat H_0$. A quench from $\hat H_f$ to $\hat M(t_\star)$ therefore freezes the state reached at any chosen time $t_\star$. More generally, the unitary conjugation in Eq.~\eqref{eq:emergent-hamiltonian} preserves every eigenvalue and its degeneracy, so that the full storage spectrum is inherited from $\hat H_0$. When one prepares a tensor product of Bell pairs between mirror-symmetric sites of a chain [see Sec.~\ref{subsec:Bell} and Fig.~\ref{fig:concept}(a)], this construction gives an exact storage Hamiltonian containing only onsite potentials and nearest-neighbor hopping at every preparation time~\cite{SurGuoMondaini2026}. Entanglement between distant sites can thus be prepared and frozen without couplings between them, making the protocol an interesting candidate for experimental emulation in platforms with limited connectivity, such as superconducting quantum circuits, where engineered, site-dependent nearest-neighbor couplings have already been used for quantum state transfer in chains~\cite{LiEtAl2018} and in two-dimensional networks~\cite{XiangEtAl2024}.

Nonetheless, practical applications must account for environmental effects, since a bath can drive the stored eigenstate to decay into other states. In Ref.~\cite{SurGuoMondaini2026}, we described this decay with relaxation and dephasing rates fixed by the coherence times $T_1$ and $T_2$, which do not depend on the energies of the transitions they induce. In general, however, the operator that couples to the bath determines which transitions are possible, while their energy differences determine the noise strength they encounter~\cite{SchoelkopfEtAl2003,ClerkEtAl2010}. To lowest order in the coupling, the rate of each transition is the squared matrix element of the coupling operator multiplied by the noise spectrum at the corresponding Bohr frequency, as we illustrate in Fig.~\ref{fig:concept}(c) [and later show in Sec.~\ref{subsec:environment}]. This frequency dependence offers another use for the emergent-Hamiltonian protocol.

Since the storage spectrum is inherited from $\hat H_0$, different reference Hamiltonians sharing the initial state as an eigenstate freeze the same prepared state while placing its transitions at different frequencies. In particular, we noted in Ref.~\cite{SurGuoMondaini2026} that the gaps inherited from $\hat H_0$ can be tuned to suppress coherent perturbations of the stored state. Here we show that the same freedom also controls its incoherent decay when the noise depends on frequency.

Realistic noise spectra are indeed far from flat. Low-frequency $1/f$ noise is ubiquitous in solid-state qubits~\cite{PaladinoEtAl2014}. In superconducting circuits, it is a dominant source of dephasing~\cite{YoshiharaEtAl2006,Bylander2011}, and classical fluctuating fields are widely used to describe it~\cite{Cywinski2008}. At higher frequencies, charge-qubit measurements also reveal a noise spectrum growing approximately linearly with frequency~\cite{AstafievEtAl2004}. Changing a transition energy can therefore increase or decrease its rate, even when the state and the environment are left fixed. The same principle underlies dynamical decoupling~\cite{ViolaKnillLloyd1999,Cywinski2008}, the control of decay by modulating the system-bath coupling~\cite{KofmanKurizki2001}, and noise spectroscopy with driven qubits~\cite{Bylander2011,YanEtAl2013}. Moving a transition toward a stronger part of the spectrum can likewise accelerate decay, as in the anti-Zeno effect~\cite{KofmanKurizki2000}. In those schemes, time-dependent control sets the frequency window the qubits sample. Here, the storage Hamiltonian sets it statically, and the shift acts on a many-body entangled state without changing it.

In what follows, we assume ideal coherent preparation and examine this freedom during storage. In Sec.~\ref{sec:memory}, we review how the Bell pairs are prepared and frozen, introduce the rescaled storage Hamiltonian $\alpha\hat M(t_\star)$, which multiplies every Bohr frequency by $\alpha$, and show that local density noise, which acts on qubits as local dephasing, cannot connect the Bell target to its degenerate partners, so that rescaling moves every transition it drives [Fig.~\ref{fig:concept}(b)]. Section~\ref{sec:forward} shows that, under $1/f$ noise, the storage time then grows in proportion to $\alpha$, a result we confirm by evolving the memory under explicit classical $1/f$ fields, whereas an Ohmic bath favors a different scale. Particle loss, which we treat for fermions as a solvable model rather than as a description of qubit relaxation, changes the particle number and is therefore sensitive to an energy shift $\mu$ per particle, which moves a zero-frequency loss transition away from zero [Fig.~\ref{fig:concept}(b)]. We also show that, among number-conserving quadratic reference Hamiltonians, $\alpha$ and $\mu$ are the only controls compatible with nearest-neighbor storage along the whole preparation. Section~\ref{sec:conclusion} summarizes our findings and contrasts the Bell memory with the GHZ memory of Ref.~\cite{SurGuoMondaini2026}. Technical details are compiled in the Appendixes, and throughout this work we set $\hbar=k_{\rm B}=1$.

\section{Bell-pair storage and noise-induced transitions}
\label{sec:memory}

\subsection{Preparing and freezing the Bell pairs} \label{subsec:Bell}

Consider an open chain of $L$ sites, with $L$ even, occupied by hard-core bosons, where the operator $\hat a_\ell$ annihilates a boson at site $\ell=0,\ldots,L-1$ and $\hat n_\ell=\hat a_\ell^\dagger\hat a_\ell^{\phantom{\dagger}}$ is the local occupation. Through the standard mapping between hard-core bosons and spin-1/2 operators~\cite{MatsubaraMatsuda1956}, each site can be identified with a qubit. As in Ref.~\cite{SurGuoMondaini2026}, we prepare the entangled state with
\begin{align}
  \hat H_f&=\sum_{\ell=0}^{L-2}J_\ell
  \left(\hat a_\ell^\dagger\hat a_{\ell+1}^{\phantom{\dagger}}
       +\hat a_{\ell+1}^\dagger\hat a_\ell^{\phantom{\dagger}}\right),
  \label{eq:Hf}\\
  J_\ell&=\frac12\sqrt{(\ell+1)(L-\ell-1)}.
  \nonumber
\end{align}
The Jordan--Wigner transformation~\cite{JordanWigner1928,LiebSchultzMattis1961} maps this one-dimensional model to free spinless fermions with annihilation operators $\hat c_\ell$, leaving the occupations $\hat n_\ell$ unchanged. In the one-particle sector, the hopping matrix has exactly the matrix elements of $\hat S_x$ for a spin $s=(L-1)/2$, with site $\ell$ corresponding to $S_z=\ell-s$, so that each fermion evolves as a large spin precessing about the $x$ axis. The same engineered hopping underlies perfect state transfer~\cite{ChristandlEtAl2004,AlbaneseEtAl2004,Kay2010}.

The initial state is the half-filled density wave $\ket{\mathrm{DW}}=\ket{1010\cdots10}$, which is an eigenstate of the reference Hamiltonian
\begin{equation}
  \hat H_0=\sum_{\ell=0}^{L-1}\ell\hat n_\ell.
  \label{eq:H0}
\end{equation}
Its time evolution under $\hat H_f$ can therefore be frozen at any time using Eq.~\eqref{eq:emergent-hamiltonian}. In particular, at $t_\star=\pi/2$ the state becomes, up to an overall phase,
\begin{equation}
  \ket{\psi_\star}
  =\bigotimes_{\ell=0}^{L/2-1}
    \frac{\ket{10}_{\ell,L-1-\ell}
          -\ii\ket{01}_{\ell,L-1-\ell}}{\sqrt2},
  \label{eq:bell-product}
\end{equation}
a product of maximally entangled pairs between mirror sites~\cite{AlkurtassBanchiBose2014,SurGuoMondaini2026}. The same concentric pairing, up to the phases within each pair, becomes the ground state of nearest-neighbor chains whose couplings decay rapidly away from the center, where it was introduced as an example of volume-law entanglement~\cite{VitaglianoRieraLatorre2010} and is known as the rainbow state~\cite{RamirezRodriguezLagunaSierra2015}. Here, it is instead reached dynamically and stored as an excited eigenstate. Quenching to $\hat M(t_\star)$ freezes these pairs, and the closed form in Appendix~\ref{app:conventions} shows that this step requires only onsite terms and nearest-neighbor hopping, even though the pairs extend across the whole chain.

Since the conjugation in Eq.~\eqref{eq:emergent-hamiltonian} fixes the eigenstates of the storage Hamiltonian but not its energy scale, we are free to use
\begin{equation}
  \hat M_\alpha(t_\star)=\alpha\hat M(t_\star),\qquad \alpha>0,
  \label{eq:scaled-storage}
\end{equation}
which leaves the Bell state and every other storage eigenstate unchanged while multiplying all energy differences by $\alpha$. A second control, the energy shift $\mu$ per particle that also appears in Fig.~\ref{fig:concept}, leaves every transition that conserves the particle number unchanged, and we therefore introduce it only in Sec.~\ref{subsec:loss}, where particle loss makes it necessary.

\begin{figure*}[t]
  \centering
  \figtwo{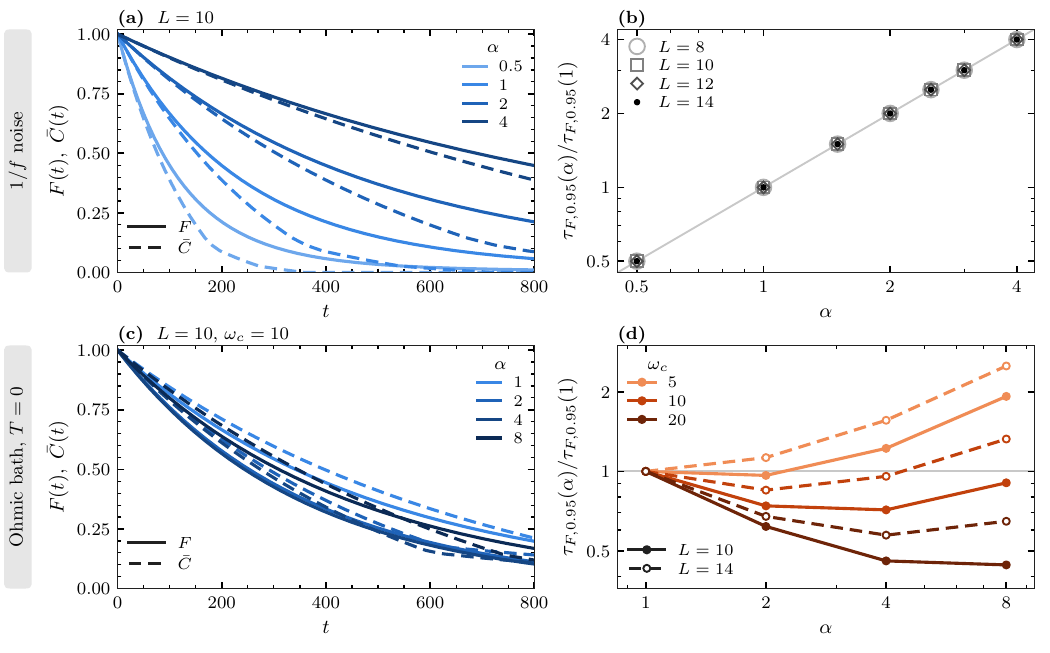}{figures/figure2_v2.pdf}
\caption{Bell memory under local density noise. In (a,c), solid curves show the target fidelity $F$ and dashed curves the mean mirror-pair concurrence $\overline C$. For symmetric $1/f$ noise, (a) shows $L=10$ at $\alpha=0.5,1,2,4$, and (b) shows the times to reach $F=0.95$, normalized to $\alpha=1$, at $L=8,10,12,14$. The gray line is the guide $y=\alpha$. For the zero-temperature Ohmic bath, (c) shows $L=10$ at $\omega_c=10$ and $\alpha=1,2,4,8$, and (d) shows lifetime ratios at $\omega_c=5,10,20$ for $L=10$ (solid) and $14$ (dashed). The $1/f$ parameters are $\eta=1$, $\omega_{\rm IR}=10^{-4}$ and $\kappa=\pi\,10^{-3}$, and the Ohmic coupling is $g_{\rm O}^{\,2}=10^{-3}$.}
  \label{fig:bell-forward}
\end{figure*}

\begin{figure*}[t]
  \centering
  \figtwo{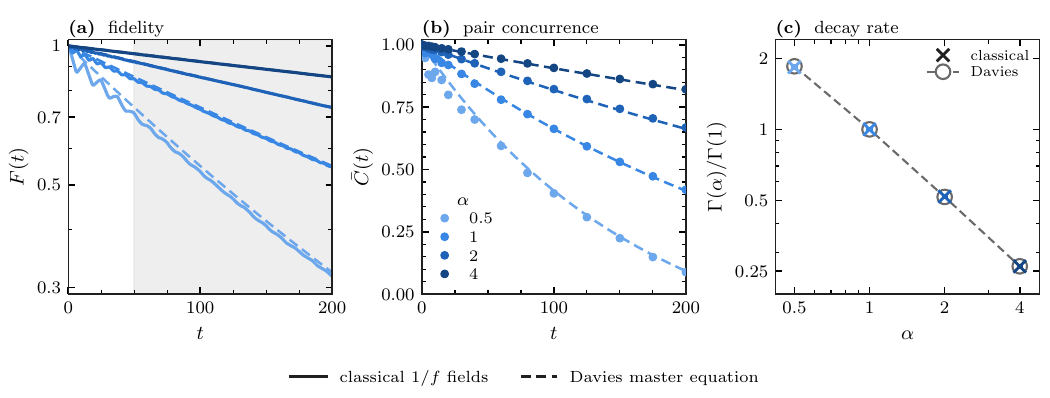}{figures/figure3_v2.pdf}
\caption{Bell storage under classical $1/f$ fields and the Davies master equation. Results are for $L=6$ and $\alpha=0.5,1,2,4$, and the classical means use $4000$ independent realizations per $\alpha$. (a) Target fidelity on a logarithmic vertical axis: solid curves with standard-error bands show the classical-field means, and dashed curves show the Davies results. The fitting interval $50\leq t\leq200$ is shaded. (b) Mean mirror-pair concurrence: markers show the classical results with bootstrap standard-error bars, and dashed curves show the Davies results. The two-site density matrices are averaged over realizations before concurrence is calculated. (c) Decay rates $\Gamma(\alpha)$ from straight-line fits to $-\ln F(t)$ over the shaded interval in (a), normalized to the rate at $\alpha=1$ in each description. Crosses with bootstrap standard-error bars show classical fields, and open circles show Davies dynamics. The field parameters are $A_\epsilon^2=10^{-3}$, corresponding to $\kappa=\pi\,10^{-3}$ as in Fig.~\ref{fig:bell-forward}, $\eta=1$, $\omega_{\rm IR}=10^{-4}$, and an upper frequency cutoff of $32$.}
  \label{fig:stochastic}
\end{figure*}

\subsection{Coupling to the environment}
\label{subsec:environment}

To see how the storage scale $\alpha$ affects the memory, we couple each local density to an independent, identical environment [Fig.~\ref{fig:concept}(a)]. Since $\hat n_\ell=(1+\hat\sigma_\ell^z)/2$, this corresponds to local dephasing noise in the qubit language, although in the basis of storage eigenstates it mostly drives transitions at nonzero frequencies. We describe the resulting dynamics by the Davies master equation~\cite{Davies1974,Gorini1976,Lindblad1976},
\begin{align}
  \frac{\dd\hat\rho}{\dd t}
  ={}&-\ii[\hat M_\alpha(t_\star),\hat\rho]
  +\sum_{\ell,\omega}S(\omega)
    \mathcal D[\hat n_\ell(\omega)]\hat\rho,\nonumber\\
  \mathcal D[\hat O]\hat\rho
  :={}&\hat O\hat\rho\hat O^\dagger
  -\tfrac12\{\hat O^\dagger\hat O,\hat\rho\},
  \label{eq:davies}
\end{align}
where $t$ now measures the time elapsed after the quench, with $\hat\rho(0)=\ket{\psi_\star}\bra{\psi_\star}$. The operator $\hat n_\ell(\omega)$ collects all density transitions at site $\ell$ that release energy $\omega$, and $S(\omega)$ is the corresponding bath rate. A positive (negative) frequency denotes energy released (absorbed) by the memory. This description retains the energy dependence of the bath while allowing us to follow the full density matrix, and the frequency components and the assumptions behind the master equation are detailed in Appendix~\ref{app:open-system}.

Let $E_\star$ and $E_f$ be the energies of the target and of an orthogonal storage eigenstate $\ket f$, separated by the transition frequency $\Omega_f=E_\star-E_f$. Equation~\eqref{eq:davies} then gives the initial rate at which $\ket f$ is populated,
\begin{equation}
  \left.\frac{\dd}{\dd t}\bra f\hat\rho(t)\ket f\right|_{t=0}
  =S(\Omega_f)\sum_\ell|\bra f\hat n_\ell\ket{\psi_\star}|^2.
  \label{eq:generic-rate-map}
\end{equation}
The energy difference thus selects a frequency in the bath spectrum, while the matrix elements determine how strongly the bath couples the two states. Figure~\ref{fig:concept}(c) groups these matrix elements by frequency and shows that changing $\alpha$ moves the frequencies while leaving the matrix elements fixed. Rescaling cannot, however, move a transition between degenerate states away from $\Omega_f=0$.

\subsection{Density noise and degenerate partners}
\label{subsec:selection}

An example of that is the Bell target $|\psi_\star\rangle$, which has orthogonal states at the same many-body energy within its particle-number sector for $L\geq6$. Whether these states cause direct decay depends on the density matrix elements. The one-particle eigenstates of $\hat M_\alpha(t_\star)$, which we call storage orbitals, make those elements simple to understand. Orbital $p=0,\ldots,L-1$ is obtained by evolving site state $\ket p$ under $\hat U(t_\star)$ and has energy $\alpha p$. Let $\hat d_p^\dagger$ create a fermion in this orbital. The density wave initially occupies the even sites, so the Bell target occupies the even-labeled storage orbitals [Fig.~\ref{fig:concept}(b)].

A local density $\hat n_\ell=\hat c_\ell^\dagger\hat c_\ell^{\phantom{\dagger}}$ is a number-conserving quadratic operator. More generally, any $\hat A=\sum_{ij}h_{ij}\hat c_i^\dagger\hat c_j^{\phantom{\dagger}}$, with arbitrary coefficients $h_{ij}$, can move only one fermion between orbitals. Apart from its component along the target itself, $\hat A\ket{\psi_\star}$ therefore contains only states obtained by replacing an occupied even orbital $p$ with an empty odd orbital $j$, and the energy released (or absorbed) in this substitution is $\Omega_f=\alpha(p-j)$. Since $p-j$ is an odd integer, it cannot vanish, and its smallest possible magnitude is one. Thus $|\Omega_f|\geq\alpha$ for every such transition. As a result, a different state with the same particle number and the same energy must change at least two occupied orbitals, which a single quadratic operator cannot do. Consequently,
\begin{equation}
  \bra m\hat A\ket{\psi_\star}=0
  \quad\text{for}\quad E_m=E_\star,
  \quad\text{if}\quad\ket m\perp\ket{\psi_\star}.
  \label{eq:zero-frequency-selection}
\end{equation}
The target therefore has no direct zero-frequency decay under local density noise. Rescaling $\alpha$ is thus sufficient to move every density transition that takes it to a different state, despite the many-body degeneracy.

\section{Tuning the storage spectrum of the Bell memory}
\label{sec:forward}

\subsection{Density noise with \texorpdfstring{$1/f$}{1/f} and Ohmic spectra}
\label{subsec:density}

Since every density transition out of the Bell target can be moved by rescaling, we now keep the preparation time at $t_\star=\pi/2$ and examine how the storage scale $\alpha$, which in practice is bounded by the largest couplings a given device can reach, affects the memory under two bath spectra, summarized in Fig.~\ref{fig:bell-forward}. We first consider a symmetric bath rate of strength $\kappa$ that decreases as $1/|\omega|^\eta$, with exponent $\eta>0$, above a low-frequency cutoff $\omega_{\rm IR}$,
\begin{equation}
  S_{1/f}(\omega)=
  \frac{\kappa}{\max(|\omega|,\omega_{\rm IR})^\eta}.
  \label{eq:one-over-f}
\end{equation}
As indicated by Eq.~\eqref{eq:zero-frequency-selection}, this rate, although largest near zero frequency, cannot drive the Bell target into a degenerate partner, and the remaining transitions, which occur at $|\Omega_f|\geq\alpha$, encounter weaker noise as $\alpha$ increases.

We follow both the complete stored state, through the target fidelity $F(t)=\bra{\psi_\star}\hat\rho(t)\ket{\psi_\star}$, and the entanglement within its mirror pairs, through the mean concurrence $\overline C$ defined in Appendix~\ref{app:concurrence}, which averages the two-qubit entanglement over the pairs. For $\eta=1$, used in all $1/f$ data shown here, both quantities decay more slowly as $\alpha$ is increased [Fig.~\ref{fig:bell-forward}(a)]. The time $\tau_{F,0.95}$ at which the fidelity reaches $F=0.95$ grows in proportion to $\alpha$ for all sizes in Fig.~\ref{fig:bell-forward}(b), consistent with the scaling $S_{1/f}(\alpha\omega)=\alpha^{-\eta}S_{1/f}(\omega)$ obeyed by every rate out of the target, whose transitions all lie far above $\omega_{\rm IR}$. The details of the density-noise calculation are given in Appendix~\ref{app:bell-davies-method}.

Increasing an energy scale need not always improve storage, however, and to illustrate how the bath spectrum determines the outcome, we also consider a zero-temperature Ohmic rate,
\begin{equation}
  S_{\rm O}(\omega)
  =g_{\rm O}^{\,2}\omega e^{-\omega/\omega_c}\Theta(\omega),
  \label{eq:ohmic}
\end{equation}
where $g_{\rm O}^{\,2}$ sets the coupling strength, $\omega_c$ is the cutoff, and $\Theta$ is the Heaviside function. At zero temperature, the bath can absorb energy released by the memory but cannot supply energy to it, and its rate grows with frequency up to $\omega_c$ and decays beyond it. Moving the Bell transitions through this spectrum can therefore either accelerate or suppress the decay [Fig.~\ref{fig:bell-forward}(c)]. Below the cutoff, increasing $\alpha$ pushes the transitions toward stronger noise and accelerates the decay, so that storage improves only once most of them lie beyond $\omega_c$. For $\omega_c=20$, the lifetimes at all displayed $\alpha>1$ remain shorter than at $\alpha=1$, whereas for the lower cutoff $\omega_c=5$ the largest scales shown yield longer storage [Fig.~\ref{fig:bell-forward}(d)]. The same Bell state can thus benefit from different storage scales, depending on the spectrum of the bath.

\begin{figure*}[t]
  \centering
  \figtwo{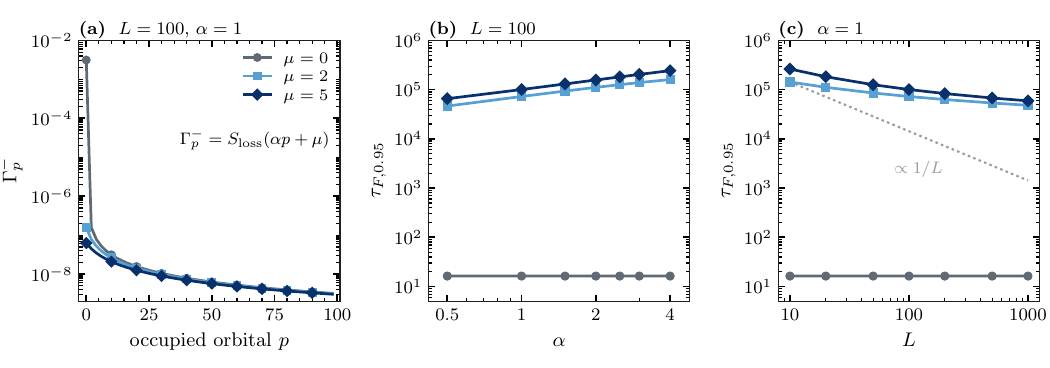}{figures/figure4_v2.pdf}
\caption{Bell memory under fermionic loss. (a) Removal rates $\Gamma_p^-$ of the occupied (even) storage orbitals at $L=100$ and $\alpha=1$, for $\mu=0,2,5$. (b) Time $\tau_{F,0.95}$ to reach $F=0.95$ versus $\alpha$ at $L=100$. (c) The same time versus $L$ at $\alpha=1$. The dotted line shows the $1/L$ decrease expected for a frequency-independent removal rate. The removal rate is $S_{\rm loss}(\omega)=\kappa/\max(|\omega|,\omega_{\rm IR})^\eta$, with $\eta=1$, $\omega_{\rm IR}=10^{-4}$ and $\kappa=\pi\,10^{-7}$. Since all lifetimes scale as $1/\kappa$, the value of $\kappa$ only sets the unit of time.}
  \label{fig:loss}
\end{figure*}

\subsection{Storage under classical \texorpdfstring{$1/f$}{1/f} fields}
\label{subsec:classical}

We next test the storage-scale control beyond the Davies description, by evolving the memory under explicit time-dependent $1/f$ fields, a standard description of low-frequency dephasing~\cite{Cywinski2008,Bylander2011}. For each realization, the Bell state evolves under
\begin{equation}
  \hat H_\xi(t)=\hat M_\alpha(t_\star)
               +\sum_\ell\xi_\ell(t)\hat n_\ell,
  \label{eq:stochastic-H}
\end{equation}
where the real fields $\xi_\ell(t)$ at different sites are independent. Each field is built from a spectral density $\mathcal P_\xi(\nu)$, specified at positive frequencies, as a sum of cosines at sampled frequencies $\nu_q>0$ with integration weights $\Delta\nu_q$ and phases $\phi_{\ell q}$ drawn independently and uniformly from $[0,2\pi)$,
\begin{equation}
  \xi_\ell(t)=\sum_q
  \sqrt{2\mathcal P_\xi(\nu_q)\Delta\nu_q}
  \cos(\nu_q t+\phi_{\ell q}).
  \label{eq:noise-synthesis}
\end{equation}
We take $\mathcal P_\xi(\nu)=A_\epsilon^2/\nu^\eta$ with $\eta=1$ inside the sampled band, which corresponds to $\kappa=\pi A_\epsilon^2$ in Eq.~\eqref{eq:one-over-f}, so that both descriptions refer to the same noise strength. The frequency range and the evolution procedure are given in Appendix~\ref{app:classical-method}.

In the interaction picture, the matrix elements of $\hat n_\ell$ between storage eigenstates acquire phases at the Bohr frequencies, so that field components near these frequencies drive transitions most efficiently. The Davies equation averages the effect of the fluctuations into fixed transition rates at those frequencies, whereas the explicit fields retain their temporal correlations, including the slow, nearly static components that dominate $1/f$ noise.

Figure~\ref{fig:stochastic} compares the evolution averaged over $4000$ field realizations with the Davies dynamics for $L=6$. As in Sec.~\ref{subsec:density}, increasing $\alpha$ slows the loss of both fidelity and mirror-pair entanglement [Figs.~\ref{fig:stochastic}(a) and \ref{fig:stochastic}(b)]. At short times, the explicit fields produce a faster, partly oscillatory drop of the fidelity that the Davies description does not capture, and this transient shrinks as $\alpha$ increases. Beyond it, the fidelity decays approximately exponentially, appearing as nearly straight lines on the logarithmic axis of Fig.~\ref{fig:stochastic}(a), at rates that differ from the Davies prediction by less than $3\%$ for every $\alpha$. Fits over the shaded interval also give the same dependence on the storage scale in both descriptions, with rates that decrease close to, though not exactly as, $1/\alpha$ [Fig.~\ref{fig:stochastic}(c)]. The improvement under $1/f$ noise therefore persists when the memory evolves under explicit time-dependent fluctuations.

\subsection{Fermionic loss and the energy shift \texorpdfstring{$\mu$}{mu}}
\label{subsec:loss}

The density-noise results rely on the selection rule of Sec.~\ref{subsec:selection}, which keeps every transition out of the Bell target at a nonzero frequency. We now consider a channel for which this is not the case, namely linear fermionic loss with coupling operator $\hat c_\ell$ at each site. Removing a fermion from storage orbital $p$ changes the particle number and releases the orbital energy $\alpha p$ [Fig.~\ref{fig:concept}(b)]. Unlike the energy differences probed by density noise, this energy is measured from the zero of the one-particle energies, which the reference Hamiltonian of Eq.~\eqref{eq:H0} places on site $\ell=0$. Since the density wave occupies this site, the Bell target fills orbital $p=0$, and loss has a zero-frequency transition that no choice of $\alpha$ can move.

This zero follows from our choice of reference Hamiltonian, not from a property of the Bell target. The density wave is also an eigenstate of $\alpha\hat H_0+\mu\calN$ for any $\alpha$ and $\mu$, where $\calN=\sum_\ell\hat n_\ell$ is the total particle number, and since $\hat H_f$ conserves $\calN$, Eq.~\eqref{eq:emergent-hamiltonian} turns this reference Hamiltonian into
\begin{equation}
  \hat M_{\alpha,\mu}(t_\star)
  =\alpha\hat M(t_\star)+\mu\calN.
  \label{eq:storage-family}
\end{equation}
The Bell target and the storage orbitals are unchanged, while the orbital energies become $\varepsilon_p=\alpha p+\mu$. Density noise cannot detect $\mu$, since the shift cancels in the energy difference between any two states with the same particle number. Removing a fermion, however, now releases $\alpha p+\mu$, which moves the formerly zero-frequency transition away from zero. The complementary density wave $\ket{0101\cdots01}$ makes the same point. It also evolves into mirror Bell pairs but fills the odd orbitals, so that at $\mu=0$ its smallest removal energy is $\alpha$, whereas a bath able to add fermions would fill the empty orbital $p=0$ at zero frequency. In both cases, $\mu$ is what keeps the transitions that change the particle number away from zero frequency.

Since the storage Hamiltonian is quadratic and the loss is linear, the problem is exactly solvable~\cite{Prosen2008,BarthelZhang2022}, and the identical, independent baths simplify it further. Written in terms of the storage orbitals, the local loss operators combine into a single loss channel per orbital, with rate $\Gamma_p^-=S_{\rm loss}(\varepsilon_p)$ set by the removal rate of the bath at the released energy (see Appendix~\ref{app:fermion-relaxation} for details). The occupied orbitals then empty independently, and since the target survives only if all of them keep their fermion, its fidelity is
\begin{equation}
  F(t)=\exp\!\left[-t\sum_{p\ {\rm even}}\Gamma_p^-\right].
  \label{eq:clockwork}
\end{equation}
The storage energies therefore enter only through where the occupied orbitals sample the spectrum of the bath, and for a frequency-independent removal rate neither $\alpha$ nor $\mu$ would change the lifetime.

Figure~\ref{fig:loss} shows the resulting rates and lifetimes for the same decreasing form as Eq.~\eqref{eq:one-over-f}, $S_{\rm loss}(\omega)=\kappa/\max(|\omega|,\omega_{\rm IR})^\eta$ with $\eta=1$. At $\mu=0$, orbital $p=0$ is removed at zero frequency, with rate $\kappa/\omega_{\rm IR}^\eta$, which exceeds all other removal rates by about four orders of magnitude [Fig.~\ref{fig:loss}(a)]. The lifetime is then set by the infrared cutoff alone and changes with neither $\alpha$ nor $L$ [Figs.~\ref{fig:loss}(b) and \ref{fig:loss}(c)]. Any $\mu>0$ lifts this orbital, together with all the others, to a weaker part of the spectrum, after which the lifetime no longer depends on $\omega_{\rm IR}$. The gain relative to $\mu=0$ therefore grows as the cutoff is lowered, and at $L=100$, $\alpha=1$ and $\omega_{\rm IR}=10^{-4}$ it amounts to a factor of about $4.4\times10^3$ for $\mu=2$ and $6.2\times10^3$ for $\mu=5$. Once $\mu>0$, increasing $\alpha$ lengthens storage further by moving the higher orbitals to weaker noise [Fig.~\ref{fig:loss}(b)]. The dependence on system size is weak as well [Fig.~\ref{fig:loss}(c)]. Since $\sum_{p\,{\rm even}}1/(\alpha p+\mu)$ grows only logarithmically with $L$, the lifetime for $\mu=2$ decreases by a factor of three between $L=10$ and $L=1000$, whereas a frequency-independent removal rate would reduce it in proportion to $1/L$, by a factor of one hundred. All of these gains follow from changing the storage energies alone, with the preparation, the Bell target and the bath held fixed.

We stress that fermionic loss serves here as a solvable model, and its conclusions should not be transferred directly to the relaxation of qubits. In the hard-core boson, or qubit, language, local loss is described by $\hat a_\ell=\hat P_\ell\hat c_\ell$ rather than by $\hat c_\ell$ [see Eq.~\eqref{eq:jw}]. Since the Jordan--Wigner string is unitary, the total loss rate out of the target is unchanged, $\sum_\ell\|\hat a_\ell\ket{\psi_\star}\|^2 =\sum_\ell\|\hat c_\ell\ket{\psi_\star}\|^2$, but its distribution over frequencies is not. Whereas $\hat c_\ell$ removes a single storage orbital at energy $\alpha p+\mu$, the string also rearranges the remaining particles, so that $\hat a_\ell$ connects the target to many final states with different energies. As a result, the orbital solution of Eq.~\eqref{eq:clockwork} no longer holds. Since every final state still has one particle fewer than the target, $\mu$ shifts all loss frequencies by the same amount, but some of these final states lie above the target in storage energy, so that at $\mu=0$ the qubit loss spectrum extends to negative frequencies, down to $-3\alpha$ for $L=6$ and $-6\alpha$ for $L=8$. A shift that clears zero frequency for fermionic loss therefore need not clear it for qubits, and at $\alpha=1$ and $\mu=2$ about $4\%$ and $1\%$ of the qubit loss weight, for $L=6$ and $L=8$, is still at zero frequency.

In particular, in superconducting circuits a uniform term $\mu\calN$ is already present in the laboratory frame through the qubit frequency, so that relaxation samples the bath near that frequency rather than near zero. The loss results therefore illustrate how $\mu$ controls a zero-frequency channel when the coupling removes one storage orbital at a time, as for fermions, rather than predicting qubit lifetimes.

\subsection{Completeness of the local storage family}
\label{sec:local-family}

So far, we have used $\alpha$ to move the density transitions and $\mu$ to move the zero-frequency loss transition, and it is natural to ask whether other reference Hamiltonians offer further control while keeping the storage Hamiltonian local. Since $\alpha\hat H_0+\mu\calN$ has the density wave as an eigenstate for any $\alpha$ and $\mu$, the family of Eq.~\eqref{eq:storage-family} extends to every preparation time,
\begin{equation}
 \hat M_{\alpha,\mu}(t_\star)
 =\alpha\hat U(t_\star)\hat H_0\hat U^\dagger(t_\star)
   +\mu\calN.
 \label{eq:all-time-storage-family}
\end{equation}
As shown in Ref.~\cite{SurGuoMondaini2026}, the emergent Hamiltonian $\hat U(t_\star)\hat H_0\hat U^\dagger(t_\star)$ contains only onsite potentials and nearest-neighbor hopping at every preparation time [see Eq.~\eqref{eq:local-storage}], and multiplying these terms by $\alpha$ and adding $\mu\calN$ preserves this locality. A single reference Hamiltonian can therefore freeze, with local couplings, whichever state the preparation is stopped at, not only the Bell product, so that the storage step does not have to be redesigned for each target. We now show that, among number-conserving quadratic reference Hamiltonians, requiring locality at every preparation time leaves only $\alpha$ and $\mu$.

Let $\hat H_{\rm ref}$ be such a reference, and let $h_{\rm ref}$ and $h_f$ be the one-particle matrices of $\hat H_{\rm ref}$ and $\hat H_f$, with $(h_f)_{\ell,\ell+1}=(h_f)_{\ell+1,\ell}=J_\ell$. At $t=0$ the emergent Hamiltonian is the reference itself, which must therefore be nearest neighbor and have the density wave as an eigenstate. Since neighboring sites of the density wave have different occupations, any hopping term in $\hat H_{\rm ref}$ moves a particle from an occupied site to an empty neighbor and produces an occupation string that no other term can cancel, so that only site potentials survive, $\hat H_{\rm ref}=\sum_\ell v_\ell\hat n_\ell$. During the preparation, $\hat H_f$ turns these potentials into hopping. To first order in $t$, the one-particle matrix of the emergent Hamiltonian acquires the term $-\ii t[h_f,h_{\rm ref}]$, which connects only neighboring sites, with amplitudes $J_\ell(v_{\ell+1}-v_\ell)$, while to second order it acquires $-(t^2/2)[h_f,[h_f,h_{\rm ref}]]$, which also connects sites two bonds apart, with
\begin{equation}
 [h_f,[h_f,h_{\rm ref}]]_{\ell,\ell+2}
 =J_\ell J_{\ell+1}
 (v_{\ell+2}-2v_{\ell+1}+v_\ell).
 \label{eq:local-potential-constraint}
\end{equation}
Since every coupling $J_\ell$ is nonzero, locality at all times requires this second difference of the potentials to vanish, so that the potentials change by the same amount $\alpha$ across every bond. Writing $v_\ell=\mu+\alpha\ell$ then gives $\hat H_{\rm ref}=\alpha\hat H_0+\mu\calN$. Conversely, every reference of this form is local at all times by Eq.~\eqref{eq:local-storage}, so that Eq.~\eqref{eq:all-time-storage-family} is the complete family of local storage Hamiltonians, with the remaining details given in Appendix~\ref{app:local-family}.

Locality at every preparation time is a stronger condition than the Bell memory alone requires. At the Bell point, every nearest-neighbor quadratic Hamiltonian with mirror-symmetric potentials and with hopping amplitudes that change sign under the reflection $\ell\to L-1-\ell$ keeps the Bell target as an eigenstate, a family with $3L/2-1$ real parameters (Appendix~\ref{app:local-family}). Which local Hamiltonians share a given eigenstate is a question of broad interest~\cite{GarrisonGrover2018,ChertkovClark2018,QiRanard2019,GjonbalajEtAl2026}, and these additional parameters would allow more general storage spectra for the Bell memory, a possibility we do not explore here. Within the family of Eq.~\eqref{eq:all-time-storage-family}, the preparation time $t_\star$ selects the state to be stored, while $\alpha$ and $\mu$ set its storage energies without changing that state or introducing longer-range couplings.

\section{Conclusion}
\label{sec:conclusion}

Emergent Hamiltonians freeze a prepared state by making it an eigenstate, and we have shown that the freedom left in this construction can be used to adapt the memory to its environment. For the density wave evolved under the engineered chain, every state reached during the preparation can be frozen with onsite potentials and nearest-neighbor hopping, and among number-conserving quadratic references that achieve this at every preparation time, only an energy scale $\alpha$ and an energy shift $\mu$ per particle remain. For the Bell memory, a selection rule lets $\alpha$ move every transition that local density noise drives, which under $1/f$ noise lengthens storage in proportion to $\alpha$, while $\mu$ moves the zero-frequency transition of fermionic loss to weaker noise. Since both controls act on the storage energies alone, with the prepared state and the environment held fixed, they can complement other means of protecting entanglement.

The Bell target owes this behavior to its position inside a degenerate spectrum, where the selection rule, rather than an energy gap, keeps its transitions away from zero frequency. The GHZ memory of Ref.~\cite{SurGuoMondaini2026}, which we analyze in Appendix~\ref{app:ghz}, illustrates the opposite situation. There, one-axis twisting prepares the GHZ state from a spin product, and the target becomes the unique ground state of a storage Hamiltonian whose terms act on all $L$ spins, so that an equilibrium bath at zero temperature cannot excite it. At finite temperature, rescaling then suppresses thermal absorption across a gap that depends on the noise operator. Local $\hat\sigma^z$ noise connects the target only to storage eigenstates with one spin changed, at energy $\alpha$, whereas excitation exchange with the bath changes all spins and connects it to states at energies $\alpha(L-1)$ and $\alpha L$, so that larger systems keep the GHZ state for a shorter time under the first coupling and for a longer time under the second. The same spectral freedom therefore applies, but through the familiar mechanism of gap protection~\cite{BaconBrownWhaley2001,JordanFarhiShor2006} and at the cost of locality.

The structure of the storage family also has a natural reading from the perspective of Hamiltonian learning. A local Hamiltonian reconstructed from a single eigenstate is typically determined only up to its overall scale and to conserved quantities with zero variance in that state, such as the particle number~\cite{QiRanard2019,GjonbalajEtAl2026}. Our two controls play exactly these roles, with $\alpha$ setting the scale and $\mu$ multiplying the particle number, although this minimal freedom comes from requiring locality during the whole preparation rather than from the Bell target alone (Sec.~\ref{sec:local-family}). Since the robustness of such reconstructions has recently been proposed as a probe of quantum chaos and ergodicity~\cite{GjonbalajEtAl2026}, it would be interesting to understand how much freedom remains for storing states prepared by interacting dynamics, for which the emergent Hamiltonian is generally nonlocal and only approximate local forms are available~\cite{SurGuoMondaini2026}.

\begin{acknowledgments}
Numerical calculations were performed on the Carya and Sabine clusters at the University of Houston's Research Computing Data Core. This work also used ACES at Texas A\&M High Performance Research Computing through allocation PHY240046 from the Advanced Cyberinfrastructure Coordination Ecosystem: Services \& Support (ACCESS) program, which is supported by U.S. National Science Foundation grants 2138259, 2138286, 2138307, 2137603, and 2138296. R.M.~acknowledges that the research was funded in part by the Robert A.~Welch Foundation, Grant \#L-E-0001-19921203.

The data and analysis scripts supporting this work are available on Zenodo~\cite{SurMondaini2026Data}.
\end{acknowledgments}

\appendix

\section{Exact local storage Hamiltonian}
\label{app:conventions}

This appendix collects the exact results on which the storage Hamiltonian of the main text rests. Appendix~\ref{app:exact-local} derives the closed form of the emergent Hamiltonian at every preparation time, which contains only onsite potentials and nearest-neighbor hopping and which Sec.~\ref{subsec:Bell} uses to freeze the Bell product. Appendix~\ref{app:local-family} completes the proof of Sec.~\ref{sec:local-family} that $\alpha$ and $\mu$ are the only controls that preserve this locality at every preparation time, and characterizes the larger family of local Hamiltonians that share the Bell target at the Bell point alone.

\subsection{Exact local form}
\label{app:exact-local}
We first review the noiseless preparation of the Bell product and the closed form of its emergent Hamiltonian. For the open chain, the Jordan--Wigner transformation~\cite{JordanWigner1928,LiebSchultzMattis1961} is
\begin{align}
  \hat a_\ell&=\hat P_\ell\hat c_\ell,\nonumber\\
  \hat P_\ell&=\exp\!\left(\ii\pi\sum_{m<\ell}\hat n_m\right)
             =\prod_{m<\ell}(1-2\hat n_m).
  \label{eq:jw}
\end{align}
The parity strings cancel between neighboring sites, $\hat a_{\ell+1}^\dagger\hat a_\ell^{\phantom{\dagger}} =\hat c_{\ell+1}^\dagger\hat c_\ell^{\phantom{\dagger}}$, and $\hat n_\ell=\hat c_\ell^\dagger\hat c_\ell^{\phantom{\dagger}}$, which gives the free-fermion form of Eq.~\eqref{eq:Hf}.

All operators used here are quadratic, $\hat A=\sum_{\ell m}A_{\ell m}\hat c_\ell^\dagger\hat c_m^{\phantom{\dagger}}$, so that their commutators, and their conjugation by $\hat U(t)$, follow from those of the one-particle matrices $A$. We denote by $\hat S_a$ the quadratic operator built in this way from the spin-$s$ matrix $S_a$, with $s=(L-1)/2$ and site $\ell$ corresponding to $S_z=\ell-s$. The engineered couplings $J_\ell$ are the matrix elements of $S_x$ in this basis~\cite{ChristandlEtAl2004,AlbaneseEtAl2004}, so that $\hat H_f=\hat S_x$ and $\hat H_0=s\calN+\hat S_z$. Rotation about the $x$ axis then gives
\begin{equation}
  \hat U(t)\hat S_z\hat U^\dagger(t)
  =\cos t\,\hat S_z-\sin t\,\hat S_y,
  \label{eq:spin-rotation}
\end{equation}
and in the site basis the emergent Hamiltonian becomes~\cite{SurGuoMondaini2026}
\begin{align}
  \hat M(t)={}&\sum_{\ell=0}^{L-1}
    [s+(\ell-s)\cos t]\hat n_\ell\nonumber\\
  &+\ii\sin t\sum_{\ell=0}^{L-2}J_\ell
    (\hat a_{\ell+1}^\dagger\hat a_\ell^{\phantom{\dagger}}
    -\hat a_\ell^\dagger\hat a_{\ell+1}^{\phantom{\dagger}}).
  \label{eq:local-storage}
\end{align}
Only onsite potentials and nearest-neighbor (purely imaginary, yet Hermitian) hopping appear, at every preparation time. At the Bell point, $t=\pi/2$, the potential becomes uniform and the hopping reaches its largest amplitude,
\begin{equation}
  \hat M(\pi/2)=s\calN-\hat S_y.
  \label{eq:bell-parent}
\end{equation}

\subsection{Completeness of the local family}
\label{app:local-family}

Here we give the details of the argument in Sec.~\ref{sec:local-family}, for a Hermitian, number-conserving quadratic reference $\hat H_{\rm ref}=\sum_{ab}(h_{\rm ref})_{ab}\hat c_a^\dagger\hat c_b^{\phantom{\dagger}}$, with the preparation Hamiltonian of Eq.~\eqref{eq:Hf} and the initial density wave held fixed. Since conjugation by $\hat U(t)$ maps quadratic operators to quadratic operators, the emergent Hamiltonian has the coefficient matrix
\begin{align}
 u(t)h_{\rm ref}u^\dagger(t)
 ={}&h_{\rm ref}-\ii t[h_f,h_{\rm ref}]\nonumber\\
 &-\frac{t^2}{2}[h_f,[h_f,h_{\rm ref}]]+O(t^3),
 \label{eq:local-parent-expansion}
\end{align}
with $u(t)=e^{-\ii h_ft}$. If this matrix is tridiagonal for every $t$, so is each of its Taylor coefficients. For the site potentials left by the eigenstate condition, $h_{\rm ref}=\operatorname{diag}(v_0,\ldots,v_{L-1})$, the first commutator has the elements $[h_f,h_{\rm ref}]_{ab}=(h_f)_{ab}(v_b-v_a)$, which vanish unless $a$ and $b$ are neighbors. Applying the commutator once more, sites $\ell$ and $\ell+2$ are connected only through site $\ell+1$, which gives Eq.~\eqref{eq:local-potential-constraint} for $\ell=0,\ldots,L-3$. Conversely, since $[\hat H_f,\calN]=0$, the reference $\alpha\hat H_0+\mu\calN$ has the emergent Hamiltonian $\alpha\hat M(t)+\mu\calN$, which keeps the evolved density wave as an eigenstate and, by Eq.~\eqref{eq:local-storage}, remains nearest neighbor at every $t$.

Locality at a single preparation time is a weaker condition. At the Bell point, consider the one-particle operator $K=u(\pi/2)\,\Pi\,u^\dagger(\pi/2)$, with $\Pi=\operatorname{diag}[(-1)^\ell]$. Since $\Pi$ equals $+1$ on the sites occupied by the density wave and $-1$ on the empty ones, the occupied storage orbitals span the $+1$ eigenspace of $K$, and a quadratic Hamiltonian keeps the Bell target as an eigenstate if and only if its one-particle matrix $m$ commutes with $K$. Because $u(\pi/2)$ rotates $S_z$ into $-S_y$, $K$ is a rotation by $\pi$ about the $y$ axis, up to a phase, and therefore connects only mirror sites, $K_{L-1-\ell,\ell}\propto(-1)^\ell$. For a tridiagonal $m$, the condition $[m,K]=0$ requires mirror-symmetric potentials, $m_{\ell\ell}=m_{L-1-\ell,L-1-\ell}$, and hopping amplitudes that change sign under reflection, $m_{L-1-a,L-1-b}=-m_{ab}$ for neighboring $a$ and $b$. This leaves $L/2$ potentials, one complex amplitude for each of the $L/2-1$ pairs of mirror bonds, and a purely imaginary hopping across the central bond, for a total of $3L/2-1$ real parameters. The storage family of Eq.~\eqref{eq:storage-family}, with uniform potential and imaginary hopping proportional to $J_\ell$ [Eq.~\eqref{eq:bell-parent}], is the two-parameter subfamily that remains local at all preparation times.

\section{Noise models and calculations}
\label{app:open-system}

This appendix specifies the density-noise calculations behind Figs.~\ref{fig:bell-forward} and \ref{fig:stochastic}, so that each result can be reproduced. We first show how the Davies equation is built from the frequency components of the coupling operators, how it gives the transition rates of Eq.~\eqref{eq:generic-rate-map}, and why its zero-frequency part leaves the Bell target unchanged. Appendix~\ref{app:concurrence} then defines the mirror-pair concurrence, Appendix~\ref{app:bell-davies-method} gives the construction and parameters of the Davies evolution, and Appendix~\ref{app:classical-method} describes the classical $1/f$ fields that test it.

Let $\hat\Pi_E$ project onto the complete eigenspace at energy $E$ of the storage Hamiltonian. The frequency component of a coupling operator $\hat A$ is
\begin{equation}
 \hat A(\omega):=\sum_{E'-E=\omega}\hat\Pi_E\hat A\hat\Pi_{E'}.
 \label{eq:bohr-components}
\end{equation}
It lowers the system energy by $\omega$, and transitions with the same energy difference are summed before forming the dissipator $\mathcal D$ of Eq.~\eqref{eq:davies}. The Davies equation describes weakly coupled relaxation on time scales long compared with the decay of bath correlations, over which terms oscillating at differences of distinct Bohr frequencies average out~\cite{Davies1974,Trushechkin2021}. Since $1/f$ noise has slowly decaying correlations, Sec.~\ref{subsec:classical} tests this description against explicit classical fields.

For the initial target projector $\hat\rho_\star=\ket{\psi_\star}\bra{\psi_\star}$ and an orthogonal energy eigenstate $\ket f$, the Hamiltonian term of the population derivative vanishes because the target is an eigenstate, and the anticommutator terms vanish because $\bra f\psi_\star\rangle=0$. The remaining contribution is
\begin{align}
 \bra f\mathcal D[\hat n_\ell(\omega)]\hat\rho_\star\ket f
 &=|\bra f\hat n_\ell(\omega)\ket{\psi_\star}|^2\nonumber\\
 &=\delta_{\omega,\Omega_f}|\bra f\hat n_\ell\ket{\psi_\star}|^2.
 \label{eq:population-slope-derivation}
\end{align}
Summing the bath rates gives Eq.~\eqref{eq:generic-rate-map}. The density selection rule also implies
\begin{equation}
 \hat n_\ell(0)\ket{\psi_\star}
 =\bra{\psi_\star}\hat n_\ell\ket{\psi_\star}\ket{\psi_\star}.
 \label{eq:zero-block-on-target}
\end{equation}
Thus the zero-frequency density dissipator leaves the target projector unchanged.

\subsection{Mirror-pair concurrence}
\label{app:concurrence}

For each mirror pair $(\ell,L-1-\ell)$, we trace out the other sites to obtain its physical two-site density matrix $\hat\rho_{\ell,L-1-\ell}$, whose concurrence is~\cite{HillWootters1997,Wootters1998}
\begin{align}
 C_{\ell,L-1-\ell}
 &=\max\{0,\lambda_1-\lambda_2-\lambda_3-\lambda_4\},\nonumber\\
 \overline C&=\frac2L\sum_{\ell=0}^{L/2-1}C_{\ell,L-1-\ell},
 \label{eq:mean-concurrence}
\end{align}
where $\lambda_1\geq\cdots\geq\lambda_4$ are the square roots of the eigenvalues of $\hat\rho_{\ell,L-1-\ell}(\hat\sigma_y\otimes\hat\sigma_y) \hat\rho_{\ell,L-1-\ell}^*(\hat\sigma_y\otimes\hat\sigma_y)$, with the complex conjugation taken in the two-site occupation basis. For the classical fields of Appendix~\ref{app:classical-method}, the two-site matrices are averaged over realizations before each concurrence is calculated.

\subsection{Bell density noise: Davies evolution}
\label{app:bell-davies-method}

The density-noise results of Fig.~\ref{fig:bell-forward} follow from Eq.~\eqref{eq:davies} with the $1/f$ rate of Eq.~\eqref{eq:one-over-f}, for $\eta=1$, $\kappa=\pi\,10^{-3}$ and $\omega_{\rm IR}=10^{-4}$, or with the zero-temperature Ohmic rate of Eq.~\eqref{eq:ohmic}, for $g_{\rm O}^{\,2}=10^{-3}$ and the cutoffs given in the figure. The first rate allows transitions in both energy directions, whereas the second only lets the memory release energy.

Since density coupling preserves the particle number, we work in the half-filled sector of the storage-orbital basis. A basis state with $N$ particles has energy $\alpha\sum_{p\in{\rm occ}}p+\mu N$, so that at $N=L/2$ all states with the same sum of occupied labels form a complete energy eigenspace. We construct the projectors onto these eigenspaces, form each $\hat n_\ell(\omega)$ using Eq.~\eqref{eq:bohr-components}, and evolve $\hat\rho(0)=\hat\rho_\star$ under the resulting generator, which gives the fidelity and the reduced two-site density matrices. We set $\mu=0$, since it cancels from all energy differences at fixed particle number. The times $\tau_{F,0.95}$ in Fig.~\ref{fig:bell-forward}(b) are first downward crossings of $F=0.95$, interpolated between consecutive recorded times.

\subsection{Bell storage under classical fields}
\label{app:classical-method}

Each independent field realization evolves according to
\begin{align}
 \ii\frac{\dd}{\dd t}\ket{\psi_\xi(t)}
 &=\left[\hat M_\alpha(t_\star)
   +\sum_\ell\xi_\ell(t)\hat n_\ell\right]\ket{\psi_\xi(t)},
 \nonumber\\
 \hat\rho(t)&=\mathbb E_\xi
 [\ket{\psi_\xi(t)}\bra{\psi_\xi(t)}],
 \label{eq:classical-evolution-method}
\end{align}
with $\ket{\psi_\xi(0)}=\ket{\psi_\star}$. The random-phase fields of Eq.~\eqref{eq:noise-synthesis} use the spectral density
\begin{equation}
 \mathcal P_\xi(\nu)=
 \begin{cases}
 A_\epsilon^2/\nu^\eta,&\omega_{\rm IR}\leq\nu\leq32,\\
 0,&\text{otherwise},
 \end{cases}
 \label{eq:classical-spectrum-method}
\end{equation}
where $\nu>0$, $A_\epsilon^2=10^{-3}$, $\eta=1$ and $\omega_{\rm IR}=10^{-4}$. The upper limit exceeds every frequency at which density noise drives transitions for $L=6$ and $\alpha\leq4$, which is at most $5\alpha=20$. The continuum covariance of these fields is
\begin{equation}
 \mathbb E[\xi_\ell(t+s)\xi_m(t)]
 =\delta_{\ell m}\int_0^\infty\mathcal P_\xi(\nu)
   \cos(\nu s)\,\dd\nu.
 \label{eq:stochastic-covariance}
\end{equation}
and its Fourier transform is
\begin{equation}
 \int_{-\infty}^\infty e^{\ii\omega s}
 \mathbb E[\xi_\ell(t+s)\xi_\ell(t)]\,\dd s
 =\pi\mathcal P_\xi(|\omega|).
 \label{eq:fourier-bridge}
\end{equation}
Within the sampled band, these fields therefore correspond to the Davies rate of Eq.~\eqref{eq:one-over-f} with $\kappa=\pi A_\epsilon^2$.

We use logarithmic frequency bins with 40 bins per decade below $0.05$, and uniform bins of width $0.01$ from $0.05$ to $32$. The frequencies $\nu_q$ in Eq.~\eqref{eq:noise-synthesis} are the geometric midpoints of the logarithmic bins and the arithmetic midpoints of the uniform ones, with $\Delta\nu_q$ the bin width, and the phases are drawn independently for every site and frequency. Each realization is evolved with time step $\Delta t=0.02$, using half a step under the storage Hamiltonian, a full step under the onsite noise evaluated at the midpoint, and another storage half-step.

For each $\alpha=0.5,1,2,4$, we average 4000 independent $L=6$ realizations. Figure~\ref{fig:stochastic}(c) compares straight-line fits to $-\ln F(t)$ over $50\leq t\leq200$, normalizing each description by its own $\alpha=1$ rate. The classical error bars are standard errors from 4000 bootstrap resamples of whole realizations.

\section{Independent orbital dynamics for fermionic loss}
\label{app:fermion-relaxation}

This appendix derives the target fidelity of Eq.~\eqref{eq:clockwork}, used for fermionic loss in Sec.~\ref{subsec:loss} and Fig.~\ref{fig:loss}. Appendix~\ref{app:orbital-loss} shows that, for identical and independent baths, local loss acts on the stored state as a single loss channel for each storage orbital, and Appendix~\ref{app:orbital-fidelity} solves for the orbital occupations and combines them into the fidelity, also allowing for baths that add particles.

\subsection{From local loss to orbital loss}
\label{app:orbital-loss}

Let $U_{\ell p}$ be the amplitude at site $\ell$ of the normalized storage orbital $p$, which has energy $\varepsilon_p=\alpha p+\mu$. The corresponding fermion operators satisfy
\begin{align}
  \hat d_p^\dagger&=\sum_\ell U_{\ell p}\hat c_\ell^\dagger,
  &\hat c_\ell&=\sum_pU_{\ell p}\hat d_p,\nonumber\\
  \sum_\ell U_{\ell p}^*U_{\ell q}&=\delta_{pq}.
  \label{eq:storage-orbitals}
\end{align}
In this basis,
\begin{equation}
  \hat M_{\alpha,\mu}(t_\star)
  =\sum_p\varepsilon_p\hat d_p^\dagger\hat d_p^{\phantom{\dagger}},
  \qquad
  \ket{\psi_\star}=\prod_{p\ {\rm even}}\hat d_p^\dagger\ket0,
  \label{eq:storage-fock-string}
\end{equation}
where the product is ordered by increasing $p$. Since $[\hat M_{\alpha,\mu},\hat d_p]=-\varepsilon_p\hat d_p$, the component of local loss that releases energy $\omega$ is
\begin{equation}
  \hat c_\ell(\omega)
  =\sum_{p:\,\varepsilon_p=\omega}U_{\ell p}\hat d_p.
  \label{eq:fermion-bohr-block}
\end{equation}
Since the orbital energies are distinct for $\alpha\neq0$, each frequency selects one orbital, $\hat c_\ell(\varepsilon_p)=U_{\ell p}\hat d_p$, and this single operator includes every many-body transition that removes a fermion from orbital $p$. For independent sites with the same removal rate $S_-(\omega)\equiv S_{\rm loss}(\omega)$, the normalization of the orbital then gives
\begin{align}
 S_-(\varepsilon_p)\sum_\ell
 \mathcal D[U_{\ell p}\hat d_p]\hat\rho
 &=\Gamma_p^-\mathcal D[\hat d_p]\hat\rho,\nonumber\\
 \Gamma_p^-&:=S_-(\varepsilon_p).
 \label{eq:identical-local-baths}
\end{align}
If the bath can also add particles, with rate $S_+(\omega)$ for a process that releases energy $\omega$ from the memory, the rate for adding a fermion to orbital $p$ is $\Gamma_p^+:=S_+(-\varepsilon_p)$, and the master equation becomes
\begin{align}
 \frac{\dd\hat\rho}{\dd t}={}&
 -\ii[\hat M_{\alpha,\mu}(t_\star),\hat\rho]
 \nonumber\\
 &+\sum_p\Gamma_p^-\mathcal D[\hat d_p]\hat\rho
  +\sum_p\Gamma_p^+\mathcal D[\hat d_p^\dagger]\hat\rho.
 \label{eq:orbital-gksl}
\end{align}

\subsection{Occupation equations and target fidelity}
\label{app:orbital-fidelity}

Define $\hat n_p=\hat d_p^\dagger\hat d_p^{\phantom{\dagger}}$ and $n_p(t)=\operatorname{Tr}[\hat\rho(t)\hat n_p]$. The Hamiltonian in Eq.~\eqref{eq:orbital-gksl} commutes with $\hat n_p$, and so do $\hat d_q$ and $\hat d_q^\dagger$ for $q\neq p$, so that jumps in the other orbitals do not change $n_p$. For jumps in orbital $p$, the fermion algebra gives
\begin{equation}
 \mathcal D^\dagger[\hat d_p](\hat n_p)=-\hat n_p,
 \qquad
 \mathcal D^\dagger[\hat d_p^\dagger](\hat n_p)=1-\hat n_p,
 \label{eq:clock-adjoint}
\end{equation}
where the adjoint dissipator acts on an observable $\hat X$ as $\mathcal D^\dagger[\hat O](\hat X) =\hat O^\dagger\hat X\hat O-\{\hat O^\dagger\hat O,\hat X\}/2$. Taking the expectation value yields
\begin{equation}
 \dot n_p=-\Gamma_p^-n_p+\Gamma_p^+(1-n_p).
 \label{eq:occupation-clock}
\end{equation}
Thus, for $r_p:=\Gamma_p^-+\Gamma_p^+>0$,
\begin{equation}
 n_p(t)=\frac{\Gamma_p^+}{r_p}
 +\left[n_p(0)-\frac{\Gamma_p^+}{r_p}\right]e^{-r_pt}.
 \label{eq:occupation-clock-solution}
\end{equation}
An orbital with $r_p=0$ retains its initial occupation.

The target is an orbital occupation state, with $n_p^\star=1$ for even $p$ and $n_p^\star=0$ for odd $p$, and Eq.~\eqref{eq:orbital-gksl} keeps its density matrix diagonal in this basis. A loss jump changes only $n_p=1$ to $0$, at rate $\Gamma_p^-$, and a gain jump only $n_p=0$ to $1$, at rate $\Gamma_p^+$. Since neither rate depends on the occupations of the other orbitals, the initially fixed occupations remain independent random variables, and the probability of matching the target in orbital $p$ is
\begin{equation}
 q_p(t)=n_p^\star n_p(t)+(1-n_p^\star)[1-n_p(t)],
 \label{eq:orbital-match}
\end{equation}
and the target fidelity is the probability of matching in every orbital,
\begin{equation}
 F(t)=\prod_pq_p(t).
 \label{eq:clock-product-fidelity}
\end{equation}
For pure loss, with $\Gamma_p^+=0$, empty orbitals remain empty and occupied ones survive with probability $e^{-\Gamma_p^-t}$, which gives
\begin{equation}
 F(t)=\exp\!\left[-t\sum_{p\ {\rm even}}\Gamma_p^-\right],
 \qquad
 \tau_f=\frac{-\ln f}{\sum_{p\ {\rm even}}\Gamma_p^-},
 \label{eq:pure-loss-lifetime}
\end{equation}
where $\tau_f$ is the time to reach fidelity $0<f<1$. Figure~\ref{fig:loss} evaluates these expressions with $\Gamma_p^-=S_{\rm loss}(\alpha p+\mu)$ and the parameters given in its caption, with $\tau_{F,0.95}=\tau_{0.95}$.

\section{GHZ storage under thermal noise}
\label{app:ghz}

\begin{figure*}[t]
  \centering
  \figtwo{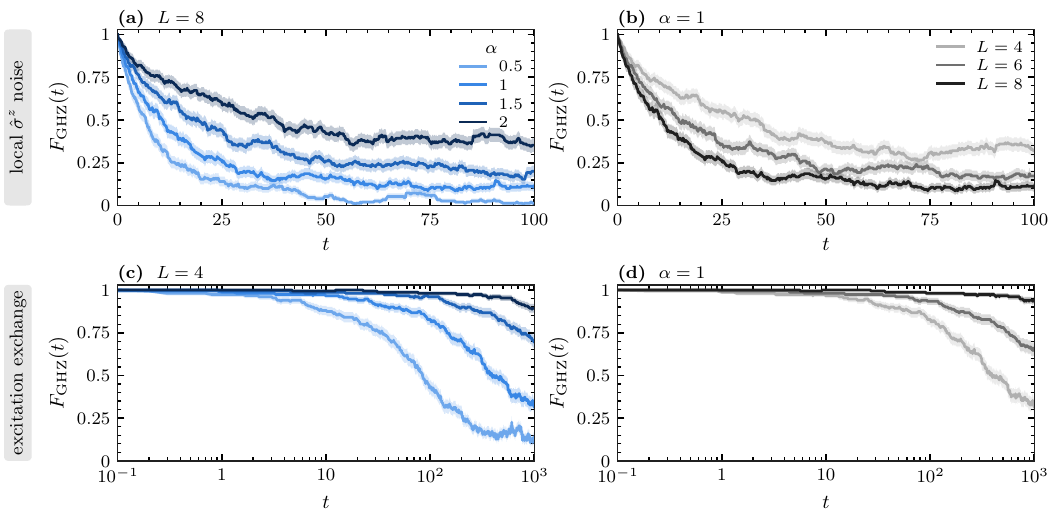}{figures/figure5_v2.pdf}
  \caption{GHZ memory under thermal noise, for local $\hat\sigma^z$ coupling (top row) and excitation exchange (bottom row). (a), (c) Target fidelity for $\alpha=0.5$, 1, 1.5 and 2, at $L=8$ for $\hat\sigma^z$ coupling and $L=4$ for exchange. (b), (d) Dependence on system size at $\alpha=1$, for $L=4$, 6 and 8. Curves are means over 150 quantum trajectories, and bands show standard errors. The independent Ohmic baths have $T=1$, $\omega_c=5$ and $g_{\rm O}^{\,2}=0.02$, and the time axis for exchange is logarithmic.}
  \label{fig:ghz}
\end{figure*}

This appendix supports the comparison between the Bell memory and the GHZ memory of Ref.~\cite{SurGuoMondaini2026} made in Sec.~\ref{sec:conclusion}. For the GHZ target, rescaling the storage Hamiltonian protects the memory through an energy gap rather than through a selection rule, and we show that the size of this gap, and hence what $\alpha$ achieves, depends on how the noise couples to the spins. Appendix~\ref{app:ghz-storage} describes the storage Hamiltonian and its spectrum, Appendix~\ref{app:ghz-noise} the two thermal couplings and the resulting fidelities in Fig.~\ref{fig:ghz}, and Appendix~\ref{app:ghz-method} the master equations and the quantum trajectories used to compute them.

\subsection{Storage Hamiltonian and spectrum}
\label{app:ghz-storage}

For $L$ spins, let $\hat{\mathcal S}_a=\frac12\sum_\ell\hat\sigma_\ell^a$ be the collective spin, with $\hat\sigma_\ell^a$ the Pauli operator at site $\ell$ along $a=x,y,z$, not to be confused with the quadratic operators $\hat S_a$ of Appendix~\ref{app:exact-local}. Starting from the product state $\ket{y_-}^{\otimes L}$, with $\ket{y_\pm}=(\ket0\pm\ii\ket1)/\sqrt2$, one-axis twisting under $\hat H_f=-\hat{\mathcal S}_z^2$~\cite{KitagawaUeda1993} reaches the GHZ state at $t_\star=\pi/2$. Taking $\hat{\mathcal S}_y$ as the reference Hamiltonian, which has this product as its unique ground state, the construction of Ref.~\cite{SurGuoMondaini2026} gives
\begin{align}
  \hat U_{\rm G}&=e^{\ii\pi\hat{\mathcal S}_z^2/2},
  &\ket{\psi_{\rm G}}&=\hat U_{\rm G}\ket{y_-}^{\otimes L},\nonumber\\
  \hat M_{\rm G}(\alpha)
  &=\alpha\hat U_{\rm G}\hat{\mathcal S}_y\hat U_{\rm G}^\dagger.
  \label{eq:ghz-storage}
\end{align}
For even $L$, the target is, up to an overall phase,
\begin{equation}
 \ket{\psi_{\rm G}}=\frac{1}{\sqrt2}
 \left[\ket{y_-}^{\otimes L}
 -\ii(-1)^{L/2}\ket{y_+}^{\otimes L}\right].
 \label{eq:ghz-explicit}
\end{equation}

The storage eigenstates follow by applying $\hat U_{\rm G}$ to the product eigenstates of $\hat{\mathcal S}_y$, and a product with $r$ spins in $\ket{y_+}$ gives an eigenstate of energy $\alpha(r-L/2)$. The GHZ target, with $r=0$, is therefore the unique ground state of $\hat M_{\rm G}(\alpha)$ for $\alpha>0$, and every spin changed relative to it costs an energy $\alpha$. Unlike the Bell storage Hamiltonian, however, $\hat M_{\rm G}(\alpha)$ is not local, since conjugation by $\hat U_{\rm G}$ attaches to each single-spin term of $\hat{\mathcal S}_y$ a string of $\hat\sigma^z$ operators on all other spins, so that every term of the storage Hamiltonian acts on all $L$ spins.

\subsection{Thermal noise and storage fidelity}
\label{app:ghz-noise}

We compare local $\hat\sigma_\ell^z$ noise with a coupling that exchanges spin excitations with the bath through $\hat\sigma_\ell^\pm=(\hat\sigma_\ell^x\pm\ii\hat\sigma_\ell^y)/2$. Since the target is the unique ground state, a bath at zero temperature cannot excite it, and we therefore take independent Ohmic baths at temperature $T$, with the spectral function $J(\omega)=g_{\rm O}^{\,2}\omega e^{-\omega/\omega_c}$ of Eq.~\eqref{eq:ohmic} for $\omega>0$ and the Bose occupation $n_{\rm B}(\omega)=(e^{\omega/T}-1)^{-1}$. Emission and absorption then occur with the rates
\begin{equation}
 S_T(\omega)=
 \begin{cases}
 J(\omega)[1+n_{\rm B}(\omega)],&\omega>0,\\
 g_{\rm O}^{\,2}T,&\omega=0,\\
 J(|\omega|)n_{\rm B}(|\omega|),&\omega<0,
 \end{cases}
 \label{eq:thermal-ohmic}
\end{equation}
where the zero-frequency value is the limit of $J(\omega)n_{\rm B}(\omega)$ as $\omega\to0^+$. As in the main text, positive $\omega$ denotes energy released by the memory, so that thermal excitation uses the rate at negative frequency.

The two couplings reach very different parts of the spectrum. Since $\hat\sigma_\ell^z$ commutes with $\hat U_{\rm G}$ and exchanges $\ket{y_+}$ and $\ket{y_-}$ at site $\ell$, it changes a single spin of the storage eigenstates and connects the target only to states at energy $\alpha$ above it. Conjugation by $\hat U_{\rm G}$ instead attaches a $\hat\sigma^z$ string to $\hat\sigma_\ell^\pm$, which changes every other spin as well, so that exchange connects the target only to states at energies $\alpha(L-1)$ and $\alpha L$. Rescaling therefore suppresses thermal absorption for both couplings, but across a gap $\alpha$ for $\hat\sigma^z$ noise and across gaps that grow with $L$ for exchange.

We follow the fidelity $F_{\rm GHZ}(t)=\bra{\psi_{\rm G}}\hat\rho(t)\ket{\psi_{\rm G}}$ for $T=1$, $\omega_c=5$ and $g_{\rm O}^{\,2}=0.02$. Increasing $\alpha$ lengthens storage for both couplings [Figs.~\ref{fig:ghz}(a) and \ref{fig:ghz}(c)], since the absorption rate $J(\Delta)n_{\rm B}(\Delta)$ decreases with the gap $\Delta$. The dependence on system size, however, is opposite for the two couplings [Figs.~\ref{fig:ghz}(b) and \ref{fig:ghz}(d)]. For $\hat\sigma^z$ noise, a larger system offers more sites through which the bath can excite the memory at the same energy, and the fidelity falls sooner, whereas for exchange the gap grows with $L$, which reduces thermal absorption and gives longer storage over the sizes and times shown. In both cases the preparation and the GHZ target remain fixed, and $\alpha$ changes only the energy that the bath must supply.

\subsection{Davies evolution and quantum trajectories}
\label{app:ghz-method}

For both couplings, the memory starts in $\ket{\psi_{\rm G}}\bra{\psi_{\rm G}}$ and evolves under $\hat M_{\rm G}(\alpha)$, with frequency components formed as in Eq.~\eqref{eq:bohr-components}. For local $\hat\sigma_\ell^z$ coupling, the Davies equation is
\begin{equation}
 \frac{\dd\hat\rho}{\dd t}
 =-\ii[\hat M_{\rm G}(\alpha),\hat\rho]
 +\sum_{\ell,\omega}S_T(\omega)
     \mathcal D[\hat\sigma_\ell^z(\omega)]\hat\rho.
 \label{eq:ghz-z-method}
\end{equation}
For excitation exchange, the components $\hat\sigma_\ell^-(\Delta)$ with $\Delta\geq0$ lower the energy of the memory as it gives an excitation to the bath, and their adjoints describe absorption, so that
\begin{align}
 \frac{\dd\hat\rho}{\dd t}={}&
 -\ii[\hat M_{\rm G}(\alpha),\hat\rho]\nonumber\\
 &+\sum_{\ell,\Delta>0}S_T(\Delta)
     \mathcal D[\hat\sigma_\ell^-(\Delta)]\hat\rho\nonumber\\
 &+\sum_{\ell,\Delta>0}S_T(-\Delta)
     \mathcal D[\hat\sigma_\ell^-(\Delta)^\dagger]\hat\rho\nonumber\\
 &+g_{\rm O}^{\,2}T\sum_\ell\left\{
     \mathcal D[\hat\sigma_\ell^-(0)]
     +\mathcal D[\hat\sigma_\ell^-(0)^\dagger]\right\}\hat\rho.
 \label{eq:ghz-exchange-method}
\end{align}
As for the Bell memory, each frequency component includes complete degenerate eigenspaces, and different sites act as independent channels.

We solve both equations with quantum trajectories, also known as the Monte Carlo wave-function method~\cite{DalibardCastinMolmer1992,DumZollerRitsch1992,MolmerCastinDalibard1993,PlenioKnight1998,Daley2014}, in which pure states evolve under a non-Hermitian effective Hamiltonian interrupted by random quantum jumps, and the average over trajectories reproduces the density matrix of the master equation. We diagonalize $\hat M_{\rm G}(\alpha)$ in the full $2^L$-dimensional spin space, take as jump operators the frequency components above, each multiplied by the square root of its rate, and draw the jump times from the decay of the norm under the effective Hamiltonian, so that the fidelity is the trajectory average of $|\bra{\psi_{\rm G}}\phi(t)\rangle|^2$. At $L=4$, these averages agree with direct evolution of the density matrix within statistical error.

\bibliography{references}

\end{document}